\documentclass[twocolumn,prl,floatfix,superscriptaddress]{revtex4}
\usepackage{graphicx}
\usepackage{natbib}
\usepackage{epstopdf}
\usepackage{color}

\begin{document}

\title{High-temperature magnetic insulating phase in ultra thin La$_{0.67}$Sr$_{0.33}$MnO$_3$ (110) films}

\author{Hans Boschker}
\affiliation{Faculty of Science and Technology and MESA$^+$ Institute for Nanotechnology, University of Twente, 7500 AE, Enschede, The Netherlands}
\author{Jaap Kautz}
\affiliation{Faculty of Science and Technology and MESA$^+$ Institute for Nanotechnology, University of Twente, 7500 AE, Enschede, The Netherlands}
\author{Evert P. Houwman}
\affiliation{Faculty of Science and Technology and MESA$^+$ Institute for Nanotechnology, University of Twente, 7500 AE, Enschede, The Netherlands}
\author{Wolter Siemons}
\affiliation{Materials Science and Technology Division, Oak Ridge National Laboratory, Oak Ridge, TN 37831, USA}
\author{Dave H. A. Blank}
\affiliation{Faculty of Science and Technology and MESA$^+$ Institute for Nanotechnology, University of Twente, 7500 AE, Enschede, The Netherlands}
\author{Mark Huijben}
\affiliation{Faculty of Science and Technology and MESA$^+$ Institute for Nanotechnology, University of Twente, 7500 AE, Enschede, The Netherlands}
\author{Gertjan Koster}
\email{g.koster@utwente.nl}
\affiliation{Faculty of Science and Technology and MESA$^+$ Institute for Nanotechnology, University of Twente, 7500 AE, Enschede, The Netherlands}
\author{Arturas Vailionis}
\affiliation{Geballe Laboratory for Advanced Materials, Stanford University, Stanford, CA 94305, USA}
\author{Guus Rijnders}
\affiliation{Faculty of Science and Technology and MESA$^+$ Institute for Nanotechnology, University of Twente, 7500 AE, Enschede, The Netherlands}
\date{\today}

\begin{abstract}
 
We present a study of the thickness dependence of magnetism and electrical conductivity in ultra thin La$_{0.67}$Sr$_{0.33}$MnO$_3$ films grown on SrTiO$_3$ (110) substrates. We found a critical thickness of 10 unit cells below which the conductivity of the films disappeared and simultaneously the Curie temperature ($T_\textrm{C}$) increased, indicating a magnetic insulating phase at room temperature. These samples have a $T_\textrm{C}$ of about 560 K with a significant saturation magnetization of 1.2 $\pm$ 0.2 $\mu_\textrm{B}$/Mn. The canted antiferromagnetic insulating phase in ultra thin films of $n<$ 10 coincides with the occurrence of a higher symmetry structural phase with a different oxygen octahedra rotation pattern. Such a strain engineered phase is an interesting candidate for an insulating tunneling barrier in room temperature spin polarized tunneling devices. 
 
\end{abstract}

\maketitle
Spin polarized tunnelling has attracted significant interest, due to the possibility to add complimentary functionality to electronic devices \cite{Meservey1994, Gregg2002}. In order to generate spin polarized currents, two approaches are being pursued. At first, spin polarized tunneling using ferromagnetic metals has been investigated \cite{Julliere1975, Moodera1995, Parkin2004, Dash2009}. Secondly, tunneling through a ferromagnetic insulator can be used for spin injection \cite{Moodera1988, Santos2004, Gajek2005, Moodera2007}. A spin polarized insulator has a different bandgap for majority and minority spin charge carriers, and therefore a difference in tunnel barrier height. This results in significant spin polarization of the tunnel current, even when the difference in bandgap is small. The most promising materials, however, have a Curie temperature ($T_\textrm{C}$) significantly lower than room temperature, 69.3 K, 16.6 K and 105 K for EuO, EuS and BiMnO$_3$ respectively\cite{Moodera2007}. Therefore, ferromagnetic insulating materials with a higher $T_\textrm{C}$ are required. 

The perovskite manganite La$_{0.67}$Sr$_{0.33}$MnO$_3$ (LSMO) is widely used in spintronics, because of its half metallicity and high $T_\textrm{C}$ of 370 K \cite{Sun1996, Odonnell2000, Bowen2003, Ogimoto2003}. This has resulted in magnetic tunnel junctions with exceptionally large tunnel magnetoresistance ratio \cite{Bowen2003}. Recently, it was shown that the $T_\textrm{C}$ of LSMO can be significantly enhanced by epitaxial strain in a carefully designed superlattice geometry \cite{Sadoc2010}. A $T_\textrm{C}$ of 650 K was achieved in the LSMO/BaTiO$_3$ (LSMO/BTO) superlattice. The saturation magnetization was rather small, which is explained with the suggestion that only the center unit cell of each 3 unit cell thick LSMO layer contributes to the magnetization. The conductivity of the superlattice was not mentioned, but generally thin layer LSMO samples are insulating \cite{Huijben2008}. Therefore LSMO is an interesting material to look for ferromagnetic insulating phases with high $T_\textrm{C}$.

Perovskite oxides are well known for their wide range of properties and the possibilities of materials engineering to enhance these properties. Next to e.g. strain engineering and interface engineering, recently research has focussed on the engineering of the oxygen octahedra rotation patterns in the perovskite thin films \cite{Rondinelli2011}. It is shown that the specific oxygen octahedra pattern, which controls the film properties, depends on the strain in the layer \cite{May2010, Rondinelli2010, Vailionis2011} and, especially at interfaces, also on the rotation pattern of the substrate \cite{Hearxiv, Borisevich2010}. A structural proximity effect is present in which the rotations of one material induce rotations in the other material. In this letter, we demonstrate that the rotation pattern of LSMO grown on SrTiO$_3$ (STO) (110)\footnote{We will use the (pseudo)cubic notation for LSMO and STO in this letter.} substrates depends on the thickness of the material. For thin films, a different rotation pattern is stabilized and results in a change of properties. We present a study of ferromagnetism and electrical conductivity in ultrathin LSMO films. We found a critical thickness of 10 unit cells (uc) below which the conductivity of the films disappeared and simultaneously the $T_\textrm{C}$ of the samples increased. The magnetic insulating phase coincides with the occurrence of a different oxygen octahedra rotation pattern, analogous to the transition between the ferromagnetic insulating and ferromagnetic metallic phases in the bulk phase diagram.

LSMO thin films were grown on oxygen annealed STO (110) substrates by pulsed laser deposition (PLD), from a stoichiometric target in an oxygen background pressure of 0.27 mbar with a laser fluence of 2 J/cm$^2$ and at a substrate temperature of 780$^\circ$C, as previously optimized for LSMO growth on STO (001) \cite{Huijben2008, Boschkerthinfilm2011}. After annealing the SrTiO$_3$ (STO) (110) substrates for 1 hour at 950$^\circ$C in a 1 bar oxygen atmosphere, smooth terraces with straight stepedges and half unit cell step height (2.7\AA) were observed, using atomic force microscopy (AFM) \footnote{See EPAPS for supplementary information}. After LSMO deposition, the films were cooled to room temperature in a 1 bar pure oxygen atmosphere. For some samples, an STO capping layer of 8 uc was grown in order to investigate the effect of possible surface reconstructions. The STO was grown using identical settings as the LSMO. The surface topology of the LSMO films was determined by AFM, showing a smooth surface with roughness at the half unit cell step height (2.7\AA) level [25].

X-ray diffraction (XRD) reciprocal space maps (RSM) were collected using a PANalytical X'Pert diffractometer in high resolution mode at Stanford Nanocharacterization Laboratory, Stanford, California. The RSMs were collected around the (332) and (33$\overline{2}$) Bragg diffraction peaks. The magnetic properties of the samples were characterized with a vibrating sample magnetometer (VSM) (Physical Properties Measurement System (PPMS) of Quantum Design). At each temperature, a full hysteresis loop between 240 and -240 kA/m ($\sim$3000 Oe) was measured and the saturation magnetization was calculated after a linear background subtraction of the diamagnetic contribution of the STO substrate. The resistivity of the samples was measured in the van der Pauw configuration \cite{vdPauw1958} (PPMS). In order to obtain ohmic contacts between the aluminum bonding wires and the LSMO layer, gold contacts were deposited on the corners of the sample by using a shadow mask.

\begin{figure}
\centering
\includegraphics*[width=8.5cm]{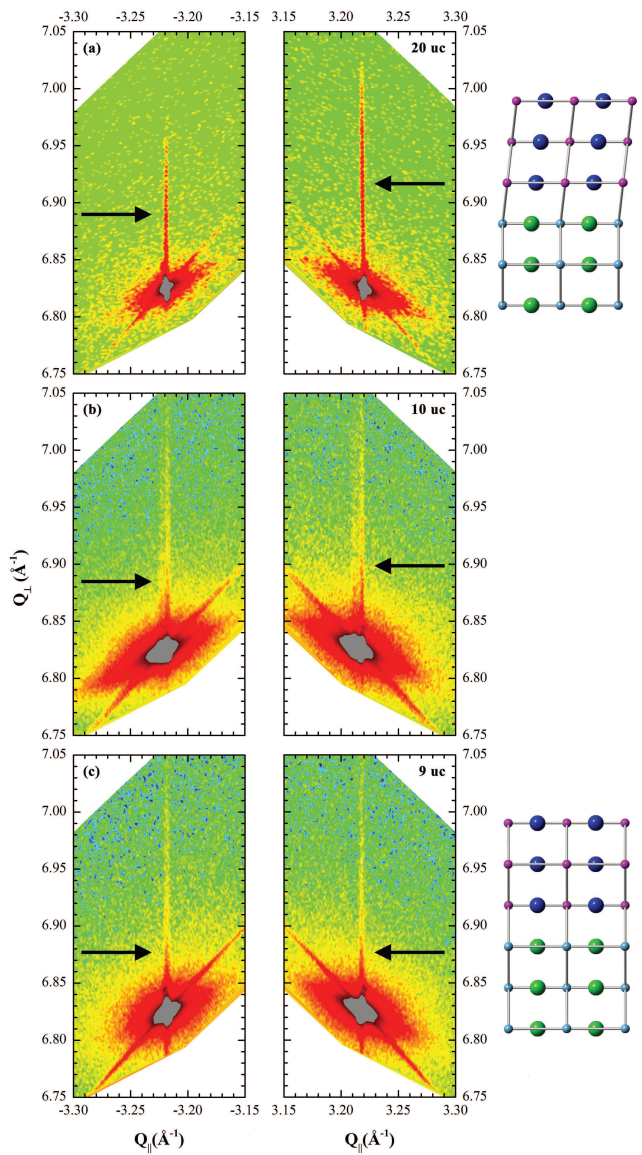}
\caption{(Color online) Reciprocal space maps of the (33$\overline{2}$) and (332) Bragg reflections of STO and LSMO of a) the 20 uc, b) the 10 uc and c) the 9 uc sample. The difference in out-of-plane momentum between the two peaks for the 20 uc sample indicates the tilting of the (001) planes, as shown with the schematics of the crystal structure (viewed along the [1$\overline{1}$0] lattice direction) on the right. The arrows indicate the positions of the film peak.}
\label{xrd}
\end{figure}

The RSMs of the 20 unit cell thick LSMO (110) layer are shown in Fig. \ref{xrd}a. It is clearly seen that (332) and (33$\overline{2}$) Bragg peak positions are different along the out-of-plane direction indicating tilting of the pseudocubic unit cell. The detailed analysis of the unit cell parameters in thicker LSMO films is presented elsewhere \cite{Boschker2010}. Most importantly, the XRD data reveal that the $\alpha$ and $\beta$ angles deviate from 90$^\circ$ and become equal to $\alpha$ = $\beta$ = 90.4 $\pm$ 0.1$^\circ$, which results in the tilt of the unit cell along the [001] direction by 0.6 $\pm$ 0.1$^\circ$. XRD data of the 9 unit cell LSMO film shown in Fig. \ref{xrd}c demonstrate no difference in the (332) and (33$\overline{2}$) Bragg peak positions. While the unit cell of this layer is distorted due to epitaxial strain, the $\alpha$ = $\beta$ = 90$^\circ$ and therefore the unit cell is not tilted. Fig. \ref{xrd}b displays the 10 unit cell thick LSMO (110) layer. The (332) and (33$\overline{2}$) Bragg peaks of the film appear at slightly different out-of-plane positions indicating onset of the orthorhombic to monoclinic structural phase transition \footnote{Note that the STO substrate of this sample displays two mosaic domains, which should not affect the properties of the film}. The n=10 sample shows relatively minor monoclinic distortion that is not completely set as seen in thick samples. Such behavior hints to the fact that n=10 sample consists of two distinct structural phases: orthorhombic and monoclinic and the LSMO(110) film at this thickness is in transition from being completely orthorhombic into being entirely monoclinic phase. The coexistence of two structural phases in LSMO thin films that are in coherence with the substrate and without the presence of misfit dislocations has been already reported previously \cite{Farag:2005go}. 

\begin{figure}
\centering
\includegraphics*[width=8cm]{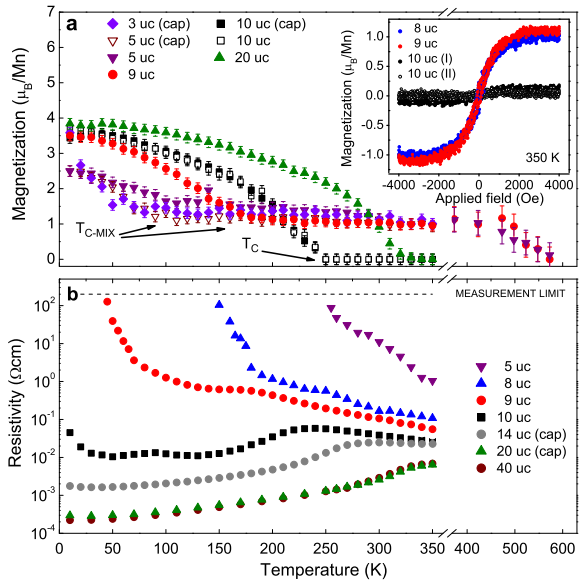}
\caption{(Color online) Temperature dependent magnetization and resistivity measurements of LSMO(110) thin films. a) Saturation magnetization of various samples with different thicknesses. The inset shows magnetic hysteresis loops of the $n=8$,  9 and 10 (I,II)  uncapped samples at $T$ = 350 K. b) Resistivity of various samples with different thicknesses.}
\label{110mag}
\end{figure}

Next, the magnetic properties of the samples are described. For thick samples, the $T_\textrm{C}$ is 350 K, equal to the $T_\textrm{C}$ of the (001) oriented samples \cite{Huijben2008}. The temperature dependence of the saturation magnetization of a set of samples with different thicknesses, as well as with and without the 8 uc STO capping layer, is presented in Fig.~\ref{110mag}a. The $n=10$ samples show the expected magnetic behaviour. The low temperature saturation magnetization is 3.6 $\pm$ 0.1 $\mu_\textrm{B}$/Mn and the $T_\textrm{C}$ is reduced to 240 K. The samples with $n < 10$ show different behaviour. At low temperatures, the saturation magnetization is high, even 2.6 $\pm$ 0.2 $\mu_\textrm{B}$/Mn for the capped $n=3$ sample. With increasing temperature, the magnetization decreases until it becomes constant at the first critical temperature, here called $T_\textrm{C-mix}$ for reasons given below. Above $T_\textrm{C-mix}$, the saturation magnetization is 1.2 $\pm$ 0.2 $\mu_\textrm{B}$/Mn. The $n < 10$ samples clearly showed a magnetic signal up to 350 K, the maximum operating temperature of the PPMS-VSM. The $n =5$ and $n =9$ samples were characterized in detail in a high-temperature VSM setup to exhibit a magnetic signal up to a $T_\textrm{C-CAFM}$ of about 560 K. To illustrate the thickness dependent magnetic behavior at 350 K, magnetic hysteresis loops of samples below ($n=8, 9$) and above ($n=10$ (two samples)) are presented in the inset of Fig.~\ref{110mag}a. The $n=10$ samples do not have a spontaneous magnetization, while the $n=8$ and 9 samples show clear hysteretic behaviour with a saturated moment of 1.1 $\pm$ 0.1 $\mu_\textrm{B}$/Mn.  

The temperature dependent resistivity measurements are presented in Fig.~\ref{110mag}b. The residual resistivity at 10 K of the thick ($n\ge20$) samples is 200 $\mu\Omega$cm, a factor of three larger than the residual resistivity of (001) oriented films \cite{Huijben2008}. This difference is most likely caused by the different crystal structure of the (110) oriented film, which changes the Mn-O-Mn orbital overlap and therefore affects the hopping integral in the double exchange mechanism. The conductivity of the $n=10$ and $n=14$ samples is reduced but still measurable at 10 K. Both samples show resistivity curves, which indicate a metal insulator transition occurs at $T_\textrm{C}$. Temperature dependent magnetoresistance measurements (not shown) support the scenario of a paramagnetic insulating/ferromagnetic metallic phase transition at $T_\textrm{C}$ in the $n\ge10$ samples. Samples with $n<10$, in contrast, show insulating behavior at all temperatures without a metal insulator transition. Taking the 9 uc LSMO(110) thin film as a representative sample, separate regions can be distinguished in the transport behavior. For high temperatures ($>$180 K) a variable range hopping model provides the best fit to the experimental data in good agreement with previous studies on transport in manganites \cite{Viret1997}. At lower temperatures there is a transition to thermally activated conductivity, which is fully developed below about 70 K.

\begin{figure}[b]
\centering
\includegraphics*[width=9cm]{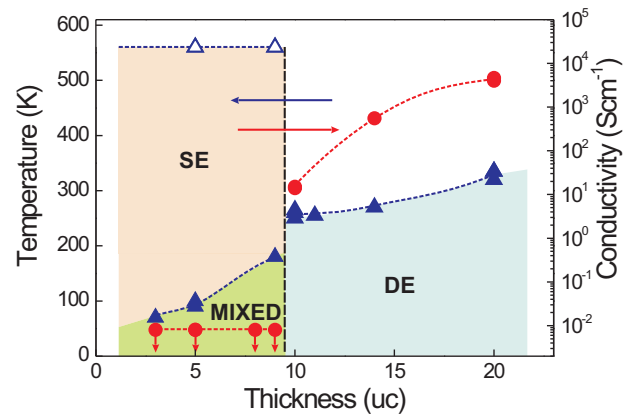}
\caption{(Color online) Constructed phase diagram showing the thickness dependence of the critical temperatures and the residual conductivity at 10 K of LSMO(110) thin films. The Curie temperatures $T_\textrm{C-CAFM}$ and $T_\textrm{C-mix}$ are indicated by open and closed triangles, respectively. The abrupt transition from 10 to 9 unit cells (uc) separates the ferromagnetic metallic (FM) state from the spin canted antiferromagnetic state (CAFM) as indicated with a vertical dashed line. The phase diagram shows the regions with double exchange (DE), super exchange (SE) and mixed interactions.}
\label{110thick}
\end{figure}

The experimental results are summarized in a phase diagram as shown in Fig.~\ref{110thick}, where the critical temperatures ($T_\textrm{C-CAFM}$ and $T_\textrm{C-mix}$) and residual (10 K) conductivity of the LSMO(110) thin films have been plotted as a function of LSMO layer thickness. The experimental data points to two distinct phases. Thin films with $n\ge 10$ are metallic and ferromagnetic, similar to the (001)$_\textrm{pc}$ oriented samples. Thin films with $n< 10$, however, are insulating with a finite spontaneous magnetization with two critical temperatures. Above $T_{\textrm{C-mix}}$, the magnetization is almost constant with temperature at 1.2 $\pm$ 0.1 $\mu_\textrm{B}$/Mn. No systematic differences between samples with and without the capping layer have been observed, indicating that the LSMO surface reconstruction does not affect the properties.

In order to explain the insulating phase with the finite spontaneous magnetization, a comparison to the bulk phase diagram, see e.g. Fujishiro \textit{et al.} \cite{Fujishiro1998}, is made. In La$_{1-x}$Sr$_x$MnO$_3$, the transition between the ferromagnetic insulating phase and the ferromagnetic metallic phase occurs at $x$=0.18, simultaneously with the orthorhombic/rhombohedral phase transition. If we first look at the structure of our samples, based on XRD measurements, films with $n\ge10$ possess a monoclinic unit cell with the space group I2/a (no.15) and can be described by Glazer's tilt system no.13: (a$^-$b$^-$b$^-$), which, due to epitaxial strain, is slightly different from the rhombohedral R-3c space group described by tilt system no.14: (a$^-$a$^-$a$^-$) \cite{Glazer1972}. As the films get thinner and reach 10 uc, the unit cell symmetry increases from monoclinic to orthorhombic. Such a change definitely affects the octahedral rotations and, at the same time deform the MnO$_6$ octahedra. These additional octahedral distortions at the interface originate from dissimilar $B$O$_6$ rotational patterns between the substrate and the coherently grown layer \cite{Hearxiv}. In this case, $B$O$_6$ rotations are absent in the STO substrate and present in the LSMO layer. In order to maintain the connectivity of the octahedra across the interface, it is anticipated that the interfacial layer will exhibit octahedral distortions that are unique compared to both substrate and layer octahedral rotation patterns. The orthorhombic to monoclinic structural phase transition at $n< 10$ is the direct consequence of such an effect. The high symmetry of the STO substrate with no octahedral rotations directly affects the structure of the LSMO layer at the interface by increasing its unit cell symmetry, which is similar to the structural phase transition of the bulk material with substitution of La by Sr. 

Subsequently the magnetic properties change due to the different Mn-O-Mn bond angles and bond lengths as compared to the thicker films. This was already reported for La$_{0.8}$Ba$_{0.2}$MnO$_{3}$ thin films and LSMO/BTO superlattices \cite{Kanki2001, Sadoc2010}. Such bond angle modification does affect the electronic orbital reconstruction at the interface and thus can result in interfacial ferromagnetic states as already has been confirmed in BiFeO$_{3}$ - La$_{0.7}$Sr$_{0.3}$MnO$_{3}$ heterostructures by x-ray magnetic circular dichroism and scanning transmission electron microscopy \cite{Yu2010,Borisevich2010}. 
The changes in bond angle result in a competition between double exchange and super exchange interactions. Such competitions was elaborated in a paper by Solovyev et al. \cite{Solovyev2001}. The authors describe a phase diagram, where an increasing super exchange interaction leads to a transition from a ferromagnetic, metallic phase to a canted antiferromagnetic, insulating phase (CAFM). They expect that there may be an intermediate mixed phase with regions, where either super exchange or double exchange interactions dominate. 

This picture is consistent with our magnetization and transport data. Interpreting the temperature and thickness dependence of magnetization and conductivity in terms of a change of the rations of the strengths of the super exchange and double exchange one can indicate in Fig.~\ref{110thick}, which interactions are dominant in the specific regions of the phase diagram. We suggest a structural coupling between LSMO layer and the STO (110) substrate is visible for this layer, where AFM dominates. In this picture decreasing temperature double exchange strength increases and a transition to a mixed phase with larger magnetization occurs. Similarly with increasing thickness there is an abrupt transition from super exchange dominated interaction to double exchange interaction, as predicted by Solovyev et al. \cite{Solovyev2001}. Another indication of the influence of the Mn-O-Mn bond angle is the observation that for the very thin films for which we expect that the octahedra rotation in the LSMO largely follows that of the substrate. The transition from AFM to a mixed phase nearly coincides with the known bulk phase transition of STO at about 100K. With thicker LSMO films the restriction of the LSMO octahedra rotations by the substrate becomes less strong.

In conclusion, due to the unique crystal structure of the $n<10$ LSMO thin films we have stabilized a canted antiferromagnetic, insulating phase with a saturation magnetization of 1.2 $\pm$ 0.2 $\mu_\textrm{B}$/Mn at a higher doping level as compared to the bulk ferromagnetic, insulating phase, which is enabled by the strong coupling to the octahedra rotations of the STO(110) substrate for the ultrathin LSMO films. The relationship between the exact microscopic structure at the LSMO(110)/STO(110) interface and the observed interfacial magnetic state is an important direction for future fundamental research on the unique octahedral tilt-controlled phenomena at the interfaces in oxide heterostructures. These strain engineered ultrathin LSMO (110) layers ( $n\le5$) show adequate material properties at room temperature, magnetism ($\sim$1.0 $\mu_\textrm{B}$/Mn) in combination with highly resistive behavior ($>$ 2 $\Omega$cm), to be an interesting candidate as spin injector for applications in spin polarized tunneling devices.

\begin{acknowledgments}
We acknowledge Lior Klein and Hans Christen for discussions and further, we wish to acknowledge the financial support of the Dutch Science Foundation (NWO) and the Dutch Nanotechnology programme NanoNed. This work is supported in part by the Department of Energy, Office of Basic Energy Sciences, Division of Materials Sciences and Engineering, under Contract No. DE-AC02-76SF00515.
\end{acknowledgments}

\bibliographystyle{apsrev}
%\bibliography{lsmo110bib}

\end{document}